\documentclass[twocolumn]{aastex631}
\usepackage{hyperref}
\usepackage{color}
\usepackage{amssymb}
\usepackage{amsmath}
\usepackage[ruled,linesnumbered]{algorithm2e}
\usepackage{footmisc}
\usepackage{ulem}

\journalinfo{arXiv}

\shorttitle{PINN RT}
\shortauthors{Chen et al.}
\begin{document}

\title{Using Physics Informed Neural Networks for Supernova Radiative Transfer Simulation}
\correspondingauthor{Lifan Wang}
\email{lifan@tamu.edu}

\author{Xingzhuo Chen}
\affiliation{George P. and Cynthia Woods Mitchell Institute for Fundamental Physics \& Astronomy, \\
Texas A. \& M. University, Department of Physics and Astronomy, 4242 TAMU, College Station, TX 77843, USA}

\author{David J. Jeffery}
\affiliation{Department of Physics \& Astronomy and Nevada Center for Astrophysics (NCfA), University of Nevada, Las Vegas, Nevada, U.S.A.}

\author{Ming Zhong}
\affiliation{Texas A. \& M. Institute of Data Science, Texas A. \& M. University. }

\author{Levi McClenny}
\affiliation{Department of Electrical and Computer Engineering, Texas A. \& M. University. }

\author{Ulisses Braga-Neto}
\affiliation{Department of Electrical and Computer Engineering, Texas A. \& M. University. }

\author{Lifan Wang}
\affiliation{George P. and Cynthia Woods Mitchell Institute for Fundamental Physics \& Astronomy, \\
Texas A. \& M. University, Department of Physics and Astronomy, 4242 TAMU, College Station, TX 77843, USA}

\begin{abstract}
    We use physics informed neural networks (PINNs) to solve the radiative transfer equation and calculate a synthetic spectrum for a Type Ia supernova (SN~Ia) SN~2011fe. 
    The calculation is based on local thermodynamic equilibrium (LTE) and 9 elements are included. 
    Physical processes included are approximate radiative equilibrium, bound-bound transitions, and the Doppler effect. 
    A PINN based gamma-ray scattering approximation is used for radioactive decay energy deposition. 
    The PINN synthetic spectrum is compared to an observed spectrum, a synthetic spectrum calculated by the Monte-Carlo radiative transfer program TARDIS, and the formal solution of the radiative transfer equation. 
    We discuss the challenges and potential of this deep-learning based radiative transfer equation solver. 
    In fact, PINNs offer the prospect of simultaneous solution of the atmosphere problem for both radiation field and thermal state throughout spacetime. 
    We have made modest steps to realizing that prospect with our calculations which required many approximations in order to be feasible at this point in the development of PINN atmosphere solutions. 
\end{abstract}

\keywords{supernovae: general, galaxies: star formation}

\section{Introduction}

Type Ia supernovae (SNe Ia) have been used as standard candles in cosmological studies \citep{Adam2021Cosmology} owing to empirical relations between the light curve properties and the maximum absolute magnitude (e.g., the Phillips relation \citep{Phillips1993DM15Relation} and Arnett rule \citep{Arnett1982Rule}). 
However, the explosion mechanism of SNe Ia is still unclear, primarily due to the computational complexity of the physical processes. 
Particularly nucleosynthesis and hydrodynamics needed in the supernova explosion simulation \citep{Gronow2021Simulation}. 

In SNe Ia, hydrodynamic and nucleosynthesis processes are only significant in the first $\sim$100 seconds. 
Thereafter the supernova ejecta expands homologously and the observed optical spectra and light curves are generated by radiative transfer and the thermal state of the ejecta. 
Therefore, simulating the radiative transfer process is necessary to estimate the density profile and element abundances of the supernova ejecta so as to put constraints on the SNe Ia explosion mechanism. 
Several well known simulation programs have been developed for the calculation of synthetic spectra for explosion model supernova ejecta structure. 
SYNOW \citep{Parrent2010SYNOW,Thomas2011SYNAPPS} uses the Sobolev method for radiative transfer calculation \citep[e.g.,][]{Rybick1978RTE} and has been used for spectral line identification. 
PHOENIX \citep{Haushildt2006PHOENIX} and CMFGEN \citep{Hillier1998CMFGEN} are more advanced simulation programs and are able to calculate 3-dimensional, non-local thermodynamic equilibrium (NLTE), time-dependent radiative transfer using the  comoving frame equation of radiative transfer \citep[e.g.,][p.~490ff]{Mihalas1978Book}. 
Programs using the Monte-Carlo method have also been developed for spectral simulation (e.g., SEDONA \citep{Kasen2006Sedona}, ARTIS \citep{Kromer2009Artis}) and have been used for spectral polarization calculations \citep{Bulla2015Polar,Livneh2022Polar}. 
In particular, TARDIS \citep{Kerzendorf2014TARDIS} is a one-dimensional radiative transfer program using the Monte-Carlo method (of radiative transfer) in which several crude approximations for NLTE effects have been implemented. 
The research reported in this paper uses the spectra from TARDIS for comparison. 

Although the aforementioned supernova spectrum simulation programs can provide results with different levels of approximation within reasonable amounts of computation time, the inverse problem, which estimates the supernova ejecta structure from an observed spectrum, still requires significant computational resources. 
In our previous study \citep{Xingzhuo2020AIAI}, a solution of the inverse problem is obtained by training a data-driven neural network on a simulated spectra data set, which contains 100,000 supernova spectra of different ejecta structure and costs $\sim$1,000,000 CPU-hours of computation time for spectral simulation and neural network training. 
Similarly, \citet{Kerzendorf2021Dalek} uses a data-driven neural network to accelerate the calculation of the forward modeling problem, and suggest the neural network could combine with the nested sampling algorithm \citep{Buchner2016Nested} to solve the inverse problem. 

Physics Informed Neural Networks (PINNs) have emerged recently as a powerful addition to traditional numerical partial differential equation (PDE) solvers \citep{Raissi2019Pinn, karniadakis2021physics}. 
The PINN approach is based on constraining the output of a deep neural network to satisfy a physical model specified by a PDE. 
Using neural networks as universal function approximators to solve PDEs had been proposed already in the $1990$'s \citep{DPT1994,lagaris1998artificial}. 
PINN capabilities at solving PDEs have been enhanced in many different ways since then by utilizing the expressive powers of deep neural networks, which are made possible by the recent advances in GPU-computing and training algorithms \citep{TF2016}, as well as computational advances in automatic differentiation methods \citep{BPRS2017}. 
A significant advantage of PINNs over traditional time-stepping PDE solvers is that PINNs are mesh-less and can solve in space and time simultaneously. 
Combined with the regression capability of deep neural networks, PINNs are also suitable for PDE-related inverse problems. 

In \cite{MMPINNRTE2021}, PINNs were applied to solve several simple monochromatic and polychromatic radiative transfer problems. 
In this paper, we employ PINNs to solve the radiative transfer equation \citep[e.g.][]{hubeny2014theory} in order to calculate an optical spectrum of Type Ia SNe SN~2011fe at 12.35 days after explosion. 

The paper is structured as follows. 
Section \ref{sec:rte} introduces the theoretical background, including the optical radiative transfer equation, the atomic physics calculation method, and the approximate gamma-ray radiative transfer calculation method. 
Section \ref{sec:pinn} describes the PINN structure used in this research and the results from the PINN calculation. 
A summary and a discussion of the future challenges for PINN-based radiative transfer calculations are given in Section \ref{sec:discuss}. 
Appendix~\ref{sec:formal} presents the formal solution of the radiative transfer equation.
The code used in this research is available on \href{https://github.com/GeronimoChen/RTPI}{https://github.com/GeronimoChen/RTPI}. 

\section{The Radiative Transfer Equation}\label{sec:rte}

In spherical symmetric coordinates, the time-independent radiative transfer equation in the rest frame is
\begin{equation}\label{eq:opticrt}
    \mathrm{cos}(\varphi)\frac{\partial I}{\partial r}-\mathrm{sin}(\varphi)\frac{\partial I}{\partial \varphi}\frac{1}{r}-j_{\rm em}\left(\frac{\bar{\nu}}{\nu}\right)^{-2}+k_{\rm abs}\left(\frac{\bar{\nu}}{\nu}\right)I=0 \,\, , 
\end{equation}
where $I$ is specific intensity (here a function of spatial coordinate, viewing direction, and frequency), 
$r$ is radius, $\varphi$ is the angle between the viewing direction and the radius vector (i.e., the viewing angle), $k_{\rm abs}$ is the comoving-frame opacity (not the rest-frame opacity), $j_{\rm em}$ is the comoving-frame emissivity (not the rest-frame emissivity), and $\left(\frac{\bar{\nu}}{\nu}\right)$ is the ratio of comoving-frame frequency to rest-frame frequency, which is given by
\begin{equation}\label{eq:Dopplershift}
    \frac{\bar{\nu}}{\nu}=\gamma[1-\mathrm{cos}(\varphi)\beta] \,\, , 
\end{equation}
where $\gamma=(1-\beta^2)^{-0.5}$ is the Lorentz factor and $\beta=v/c$ is the velocity of the material divided by the speed of light (\citealt[e.g.,][eq.~(1--3)]{Castor1972RTE};
see also \citealt[][p. 31,33,495--496]{Mihalas1978Book}). 
The formal solution of the radiative transfer equation is presented in Appendix \ref{sec:formal}. 

Note we have dropped the time dependence term because we model only the atmosphere of a SN~Ia above an inner core that provides inner boundary condition for the atmosphere. 
Time dependence in SN~Ia atmospheres has generally been found to be relatively unimportant near and even prior to maximum light \citep[e.g.,][\S~3.5]{Kasen2006Sedona} and can be neglected for our exploratory calculations. 

Because SN~Ia ejecta is expanding homologously by about the first 10 seconds after the explosion of the relatively small progenitor white dwarf \citep[e.g.,][]{Ropke2005IaSimu}, the material velocity at all observable epochs is proportional to the radius at a given time and satisfies the relation $r=v t_{\rm exp}$, where $t_{\rm exp}$ is the time since the explosion. 
Therefore, we use the material velocity to represent the spatial coordinate in the figures and elsewhere as needed. 

In the following subsections, we will introduce the different physical processes that contribute to the opacity $k_{\rm abs}$ and emissivity $j_{\rm em}$. 

\subsection{Thermal State, LTE, and Temperature Profile}

For the calculation of the thermal state of our Type~Ia supernova atmosphere model, we assume local thermodynamic equilibrium (LTE) which means that all matter occupation numbers are determined by their thermodynamic equilibrium values calculated from a single temperature. 
Imposing radiative equilibrium plus gamma-ray energy deposition (calculated as described in \S\S~\ref{subsec:gamma} and \ref{subsec:gammanet}), the LTE temperature $T$ is calculated from the quasi steady state first law of thermodynamics equation with time derivatives of energy density and adiabatic cooling omitted as being negligible which is suitable for supernova atmospheres \citep[e.g.,][\S~2.4]{Kasen2006Sedona}. 
This equation written for our model is the radiative equilibrium equation
\begin{equation}\label{eq-radiative-equilibrium-equation}
\int_{0}^{\infty}\kappa_{\nu}B_{\nu}(T)\,d\nu
=\int_{0}^{\infty}\kappa_{\nu}J_{\nu}\,d\nu
+{\cal E}_{\gamma} \,\, , 
\end{equation}
where $\nu$ is the comoving frequency, $T$ is the LTE temperature to be solved for, $\kappa_{\nu}$ is the comoving absorption opacity (implicitly evaluated at $T$), 
\begin{equation}
    B_{\nu}(T)={2h\nu^{3}\over c^{2}}
     {1\over e^{h\nu/(k_{\rm B}T)}-1}
\end{equation}
is the Planck law (with $k_{\rm B}$ being the
Boltzmann constant),
\begin{equation}
J_{\nu}=\frac{1}{4\pi}\int_{\Omega}I\, d\Omega
\end{equation}
is the comoving mean specific intensity
(with the frequency dependence of $I$ implicit
which is the convention we adopt in this paper), 
and ${\cal E}_{\gamma}$ is the (rate of) gamma-ray energy deposition per unit solid angle (e.g.., \citealt[][p.~172]{Mihalas1978Book}: see also \citealt[][\S~2.4]{Kasen2006Sedona}). 
Note
\begin{equation}
{\cal E}_{\gamma}= {E_{\gamma}\over 4\pi} \,\, , 
\end{equation}
where $E_{\gamma}$ is the (rate of) gamma-ray energy deposition introduced
in \S~\ref{subsec:gamma}. 

Using 
\begin{equation}
    B(T)={\sigma_{\rm SB}T^{4}\over\pi} \,\, ,
\end{equation}
the Planck law integrated over all frequency (with $\sigma_{\rm SB}$ being the
Stefan-Boltzmann constant),
we can rewrite Equation~(\ref{eq-radiative-equilibrium-equation}) in a form clearly implying there is a temperature to be solved for: 
\begin{equation}\label{eq-radiative-equilibrium-equation-T}
T=\left[ {\pi J\over \sigma_{\rm SB}}R_{\rm op}
       +{\pi {\cal E}_{\gamma}\over \sigma_{\rm SB} 
          \kappa_{\rm P}(T) }\right]^{1/4} \,\, ,
\end{equation}
where $J$ is the mean intensity integrated over all frequency, $\kappa_{\rm P}(T)$ is the Planck mean opacity (see definition in the equation just below), and
\begin{equation}\label{eq-opacity-ratio}
R_{\rm op}={\kappa_{J}\over\kappa_{\rm P}(T)}
  ={\int_{0}^{\infty}\kappa_{\nu}J_{\nu}\,d\nu/J
    \over
     \int_{0}^{\infty}\kappa_{\nu}B_{\nu}(T)\,d\nu/B(T)}
\end{equation}
is the opacity ratio, where $\kappa_{J}$ is the absorption mean opacity  \citep[e.g.,][p.~60]{Mihalas1978Book} and $\kappa_{\rm P}(T)$ is the aforementioned Planck mean opacity \citep[e.g.,][p.~59]{Mihalas1978Book}.
Note that the opacity ratio is independent of the scales of $J_{\nu}$, $B_{\nu}$, and $\kappa_{\nu}$. 
The independence of the scale of $J_{\nu}$ is just because $J$ (loosely speaking the total driving radiation field) appears explicitly in Equation~(\ref{eq-radiative-equilibrium-equation-T}). 
The independence of the scale of $B_{\nu}$ just follows from the definition of the Planck mean opacity. 
The independence of the scale of $\kappa_{\nu}$ can be understood by seeing that when ${\cal E}_{\gamma}=0$, the LTE radiative equilibrium temperature is independent of the scales of energy inflow and outflow to matter and these scales are the only things controlled by the scale of $\kappa_{\nu}$.
The independence of the scales of $J_{\nu}$, $B_{\nu}$, and $\kappa_{\nu}$ suggests that the opacity ratio $R_{\rm op}$ may often be of order $1$. 

Equation~(\ref{eq-radiative-equilibrium-equation-T}) is, of course, still an implicit equation for $T$. 
However, it will probably succeed as an iteration formula in most cases. 
However also, given that the opacity ratio $R_{\rm op}$ may often be of order 1 and 1 is the neutral choice given no other information, we can set it to be 1 for two characteristc temperatures derived from Equation~(\ref{eq-radiative-equilibrium-equation-T}). 
The first one, which we call the Planck law temperature (PLT), is
\begin{equation}
T_{\rm PLT}=\left( {\pi J\over \sigma_{\rm SB}}
                    \right)^{1/4} \,\, ,
\end{equation}
where ${\cal E}_{\gamma}$ is set to zero in Equation~(\ref{eq-radiative-equilibrium-equation-T}). 
The second characteristic temperature, which we call the Planck law temperature augmented (PLTA), is
\begin{equation}
\label{eq-radiative-equilibrium-equation-PLTA}
T_{\rm PLTA}=\left[ {\pi J\over \sigma_{\rm SB}}
       +{\pi {\cal E}_{\gamma}\over \sigma_{\rm SB} 
          \kappa_{\rm P}(T_{\rm PLT})}\right]^{1/4} \,\, ,
\end{equation}
where ${\cal E}_{\gamma}$ is not set to zero in Equation~(\ref{eq-radiative-equilibrium-equation-T}). 

There are two special cases where PLT is an explicit exact solution for temperature for Equation~(\ref{eq-radiative-equilibrium-equation-T}) with ${\cal E}_{\gamma}=0$. 
The two cases, of course, have $R_{\rm op}=1$ exactly. 
The first case is where the absorption opacity is grey (i.e., frequency independent, but not necessarily independent of any other variable). 
The second case is where $J_{\nu}=WB_{\nu}(T)$, where $W$ is a dilution factor and $T$ is the solution temperature itself. 
The dilution factor is 1 and the temperature $T$ at high optical depth in an atmosphere where matter collisions are sufficiently rapid to enforce mutual thermodynamic equilibrium between matter and radiation. 
Having $J_{\nu}=WB_{\nu}(T)$ with $W\neq1$ is also possible for the second special case, but perhaps rather unusually. 
Note $W<1$ may happen sometimes, but $W>1$ (i.e., a ``negative'' dilution) is probably rather rare. 
Both the first and second special cases are proven by inspection. 

Since the opacity ratio may often be of order 1, and this is the neutral approximation with no other information given, and there are two special cases where the opacity ratio is 1, PLT/PLTA is usually a good characteristic temperature for LTE and may be very good approximation in some cases. 
Of course, PLTA is not likely to be a good approximation or characteristic temperature if the second term of Equation~(\ref{eq-radiative-equilibrium-equation-PLTA}) is larger than the first term:  i.e., when gamma-ray energy deposition dominates the temperature determination. 

Note that if there are significant NLTE effects in an atmosphere, PLT/PLTA is, a~priori. as good a characteristic temperature as the one obtained by solving Equation~(\ref{eq-radiative-equilibrium-equation}) exactly since that is an LTE equation. 

Except in the special cases where PLT is exact, radiative equilibrium (i.e., energy conservation) in LTE can only be expected to be approximate for PLT/PLTA when used with a radiative transfer equation solution of radiative transfer. 
However, PTL/PLTA can be applied with conservation of energy in Monte Carlo radiative transfer with indestructible photon packets \citep{Lucy1999}. 
The use of photon packets enforces energy conservation and PTL/PLTA just becomes a rule to determine a characteristic temperature with some degree of realism for either LTE or NLTE. 

For our pioneering PINN radiative transfer calculations, we do not need high realism in the thermal state solution (which for LTE is essentially solving for the temperature profile) and in particular do not need exact radiative equilibrium (i.e., energy conservation). 
Therefore, we adopted PLTA for our temperature profile calculation and made a simplifying approximation for the Planck mean opacity $\kappa_{\rm P}(T)$: we take it to be equal to the electron scattering opacity
\begin{equation}
    k_{e}=\sigma_{\rm T} N_e \,\, , \
\end{equation}
where $\sigma_{\rm T}$ is the Thomson scattering cross section and $N_e$ is free electron number density. 
With this choice for $\kappa_{\rm P}(T)$, the PTLA formula becomes 
\begin{equation}\label{eq-PLTA-electron-scattering}
T_{\rm PLTA}=
   \left( {\pi J\over \sigma_{SB}}
   + {\pi {\cal E}_{\gamma}
   \over \sigma_{SB}\sigma_{\rm T} N_e} 
           \right)^{1/4}  \,\, . 
\end{equation}
It is, of course, formally wrong to use a scattering opacity as an absorption opacity. 
However, the usage is a calculational placeholder for a better treatment using a good Planck mean opacity from tables or a good approximate formula. 
The electron scattering opacity is probably an overestimate, and so probably minimizes in our calculations the gamma-ray energy deposition which in any case we find to be very small. 
Since the gamma-ray energy deposition is small and since we are not interested in high realism in the thermal state calculations, using electron scattering opacity for the Planck mean opacity is adequate for our purpose which as aforesaid is to solve the radiative transfer equation with PINN. 
We also solve the gamma-ray radiative transfer with PINN to show that that can be done (see \S~\ref{subsec:gammanet}). 

The temperature profile is solved for by Lambda iteration \citep[e.g.,][p.~147--150]{Mihalas1978Book}. 
We assume an initial temperature from which we calculate the bound-bound opacities (i.e., bb or line opacities) which are the only opacities we include in the radiative transfer (see \ref{Bound-Bound Transitions}): the electron scattering opacity is used only to solve for temperature from Equation~(\ref{eq-PLTA-electron-scattering}). 
The bb opacities are treated as pure absorption opacities: i.e., no line scattering is included and the lines emit thermally (see \ref{Bound-Bound Transitions}). 
Using the bb opacities and emission from the inner boundary, we calculate $J$ from the radiative transfer and then use Equation~(\ref{eq-PLTA-electron-scattering}) to calculate a new temperature profile from which new bb opacities are calculated. 
We then calculate the radiative transfer again, and so on until the temperature profile converges. 
The Lambda iteration successfully converges in our case since since the atmosphere is overall optically thin with just the bb opacities included \citep[e.g.,][p.~147--150]{Mihalas1978Book}. 

To speed up the Lambda iteration, we avoid the integrations for $J$ by using temperature (neural) network as described in \S~\ref{subsec:temperature}.

\subsection{Bound-Bound Transitions
     \label{Bound-Bound Transitions}}

The bound-bound opacity and emissivity are calculated using the local thermodynamic equilibrium (LTE) approximation. 
The spontaneous emissivity is 

\begin{equation}
    j_{\rm bb}=\frac{h}{4\pi}A_{ul}\nu_{ul} N_{i,j,u}\phi(\nu) \,\, , 
\end{equation}
where $A_{ul}$ is the Einstein $A$ coefficient for the line transition from the $u$-th level to the $l$-th level, $N_{i,j,u}$ is the number density, $i,j,u$ are the indices of element, ionization, and energy level, respectively, $\nu_{ul}$ is the spectral line frequency, and $\phi(\nu)$ is the line shape profile which satisfies the normalization condition $\int_{0}^{\infty} \phi(\nu) d\nu =1$ \citep[e.g.,][p. 78]{Mihalas1978Book}. 
The opacity corrected for stimulated emission is
\begin{equation}
    k_{\rm bb}=\frac{h}{4\pi}(N_{i,j,l}B_{lu}-N_{i,j,u}B_{ul}) \nu_{ul} \phi(\nu) \,\, , 
\end{equation}
where $B_{ul}$ and $B_{lu}$ are the Einstein $B$ coefficients for the absorption and stimulated emission processes \citep[e.g.,][p. 78--79]{Mihalas1978Book}. 
All the Einstein coefficients are downloaded from NIST spectral database. 

The realistic line profile $\phi(\nu)$ is usually the Voigt function which accounts for both natural line broadening and temperature Doppler broadening \citep[e.g.,][p. 279--281]{Mihalas1978Book}. 
These broadening effects are much smaller than the Doppler effect from the supernova ejecta velocity (typically ($\sim 10000\,\mathrm{km/s}$), and so we replace the usual realistic line profile with an artificial unrealistic one without significant error as long as it likewise has insignificant broadening. 
We choose a simple rectangular function with a 4 pixel width as the line shape profile in order to reduce the computation time. 
When the frequency grid is between $10^{14.4}\,\mathrm{Hz}$ ($12000\, \mathrm{\AA}$) and $10^{15}\,\mathrm{Hz}$ ($3000\, \mathrm{\AA}$) with 2048 sampling points uniformly sampled in the logarithmic space, the velocity resolution is $202\,\mathrm{km/s}$ and spectral line width is $808\,\mathrm{km/s}$ (which is much smaller $\sim 10000\,\ \mathrm{km/s}$). 
Note that the temperature Doppler broadening velocity in supernova ejecta (which typically have temperatures of order $10^{4}\,$K) is of order $10\,\mathrm{km/s}$ \citep[e.g.,][p. 279]{Mihalas1978Book}, and so we have introduced artificial line broadening large compared to temperature Doppler broadening, but still negligible for our calculations. 

\subsection{Level Population}

The level populations are calculated in LTE using Saha ionization equation and the Boltzmann equation. 
The relevant equations for solving for ionization state and electron density are as follows. 
First, the ratio between the two level populations (i.e., the Saha ionization equation in one version) is
\begin{equation}\label{eq:sahaboltzmann}
    \frac{N_{i,j,k}}{N_{i,j+1,0}}=N_e \frac{1}{2}\left(\frac{h^2}{2\pi m_e k_{\rm B}}\right)^{\frac{3}{2}} \frac{g_{i,j,k}}{g_{i,j+1,0}}T^{-\frac{3}{2}}e^{\frac{\chi_{i,j+1,0}-\chi_{i,j,k}}{k_{\rm B} T}} \,\, , 
\end{equation}
where $N_e$ is the electron number density, $g_{i,j,k}$ is the degeneracy factor, $\chi_{i,j,k}$ is the level energy, $i,j,k$ are as before the indices of element (which is the atomic number), ionization, and level, respectively, and 0 labels the ground state level \citep[e.g.,][p. 113]{Mihalas1978Book}. 
Second, the supernova ejecta plasma is assumed to be neutral, and so the electron number density satisfies 
\begin{equation}\label{eq:ne}
    N_e=\sum_{i} \sum_{j} j \sum_{k} N_{i,j,k} \,\, . 
\end{equation}
Third, the element abundances $N_i$ in supernova ejecta models satisfy 
\begin{equation}\label{eq:ni}
    N_{i}=\sum_{j}\sum_{k} N_{i,j,k} \,\, . 
\end{equation}
We adopt a similar algorithm as is in the TARDIS code to solve the $N_{i,j,k}$ for the above equation group (i.e., Eqs.~(\ref{eq:sahaboltzmann}), (\ref{eq:ne}), (\ref{eq:ni})). 
In the first step, we assume all the atoms are singly ionized and calculate an $N_{e,0}$. 
In the second step, the level population $N_{i,j,k,0}$ is calculated from Equation~(\ref{eq:sahaboltzmann}) and Equation~(\ref{eq:ni}) using the assumed $N_{e,0}$. 
In the third step, a newer electron density $N_{e,{\rm new}}$ is calculated from Equation~(\ref{eq:ne}) using the level population in the second step. 
In the fourth step, we calculate the mean of two electron densities: $N_{e,1}=(N_{e,0}+N_{e,{\rm new}})/2$, then go back to the second step and replace $N_{e,0}$ with $N_{e,1}$. 
We repeat the above 4 steps until there is a convergence of ionization state and electron density. 
In our calculation, the number of iteration is set to 30, which guarantees a convergent and sufficiently accurate solution in all of our tests. 

\subsection{Gamma-Ray Energy Deposition}\label{subsec:gamma}

Gamma ray photons in SNe Ia are mainly generated from the decay chain ${\rm ^{56}Ni}\rightarrow{\rm ^{56}Co}\rightarrow{\rm ^{56}Fe}$, and the photon energy are deposited as thermal energy via Compton scattering and photoelectric absorption processes. 
The pair-production process is negligible for our system \citep[e.g.,][]{Kasen2006Sedona}. 
The physical details of our gamma ray treatment (discussed below) are taken from the appendix of \citet{Kasen2006Sedona}. 
The Compton scattering opacity is
\begin{equation}
    k_{\rm C}=\sigma_{\rm T} K(x) \sum_{i} i N_i \,\, , 
\end{equation}
where $i$ is the atomic number as above, $\sigma_{\rm T}$ is the Thomson cross section for Thomson scattering \citep[e.g.,][p. 106]{Mihalas1978Book}, x is gamma ray photon energy $x={h\nu}/({m_e c^2})$, and $K(x)$ is the Klein-Nishina correction to the Thomson cross section for Compton scattering: 

\begin{widetext}
    \begin{equation}
        K(x)=\frac{3}{4}\left\{ \frac{1+x}{x^3}\left[\frac{2x(1+x)}{1+2x}-\ln(1+2x)\right]+\frac{\ln(1+2x)}{2x}-\frac{1+3x}{(1+2x)^2}\right\} \,\, . 
    \end{equation}
\end{widetext}
Note the summation is over all the bound and free electrons because the gamma-ray energies are much larger than the photoionization energies, and so all the electrons contribute to the Compton scattering effect. 

The photoelectric opacity is 
\begin{equation}
    k_{\rm p}=\sigma_{\rm T} \alpha^4\left(8\sqrt{2}\right) x^{-7/2} \sum_{i}i^5 N_i \,\, , 
\end{equation}
where $\alpha$ is fine-structure constant. 

The emissivity in the Compton scattering process is usually determined from the differential cross section $d\sigma/d\Omega$ in most of the Monte-Carlo based radiative transfer programs \citep[e.g.,][]{Kasen2006Sedona}. 
The formula for $d\sigma/d\Omega$ is
\begin{equation}\label{eq:comptonphase}
    \frac{d\sigma}{d\Omega}=\frac{3\sigma_{\rm T}}{16\pi}f(x,\Theta)^2[f(x,\Theta)+f(x,\Theta)^{-1}-\sin^2\Theta] \,\, , 
\end{equation}
where $\Theta$ is the angle between the incoming and outgoing gamma ray photon and $f(x,\Theta)$ is the energy ratio between incoming and outgoing photon and is given by 
\begin{equation}
    f(x,\Theta)=\frac{E_{\rm out}}{E_{\rm in}}=\frac{1}{1+x(1-\mathrm{cos}\Theta)} \,\, . 
\end{equation}
The average energy lost in an interaction is 
\begin{equation}\label{eq:gammalost}
    F(x)=1-\frac{1}{4\pi}\int_\Omega f(x,\Theta)\,d\Omega \,\, . 
\end{equation}

We use Equation~(\ref{eq:gammalost}) to calculate a sequence of discrete gamma ray photon energy bins which correspond to 0, 1, 2, $\dots$  6 times scattered photons. 
We then use PINN to model the number of photons over these discrete energy bins as a function of spatial coordinate $r$ and viewing angle $\varphi$. 
The ($m+1$)-th emissivity is calculated from the integral of the absorbed photon energy in the $m$-th gamma ray photon energy bin and averaged over solid angle and is written as
\begin{equation}
    j_{{\rm C},m+1}=\frac{1}{4\pi} \int_{\Omega} k_{{\rm C},m} I_{m}\, d\Omega \,\, , 
\end{equation}
where $m$ is the index of gamma energy bin. 

The time independent gamma-ray radiative transfer equation is 
\begin{equation}\label{eq:gammart}
    \mathrm{cos}(\varphi)\frac{\partial I_{m}}{\partial r}-\mathrm{sin}(\varphi)\frac{\partial I_{m}}{\partial \varphi}\frac{1}{r}+(k_{\rm C}+k_{\rm p})I_{m}-j_{\rm C ,m}-j_{\rm r}=0 \,\, , 
\end{equation}
where $m$ is the label of our 7 discrete gamma-ray energy bins (listed in Section \ref{subsec:gammanet}) and $j_{\rm r}$ is the gamma ray source in the supernova atmosphere. 
The detailed calculation procedure is discussed in Section \ref{subsec:gammanet}. 
We assume the energy lost in gamma-ray Compton scattering and photoelectric absorption processes are deposited as thermal energy locally and, as aforesaid, neglect the gamma-ray photon pair-production process. 
Therefore the gamma-ray energy deposition is 
\begin{equation}
    E_{\gamma}=\sum_m h\nu_m \int_\Omega \left[(k_{\rm C,m}+k_{\rm p})I_{m} -j_{\rm C,m}\right]\, d\Omega \,\, , 
\end{equation}
where the summation is over all the allowed gamma-ray energy bins and we count $j_{\rm C,m}$ as a negative energy deposition. 
The energy deposition $E_{\gamma}$ is used in Equation~(\ref{eq-PLTA-electron-scattering}) to calculate the plasma temperature. 

\section{The Physics Informed Neural Network}\label{sec:pinn}

The concept of solving partial differential equations (PDEs) using neural networks has a long history. 
The idea is commonly credited to \cite{lagaris1998artificial} though there is related work dating back to the late 1980's (see \cite{viana2021survey} for a historical review). 
Recently \cite{Raissi2019Pinn} proposed Physics Informed Neural Network (PINN), a modern deep neural network approach to solve forward and inverse PDE-constrained problems. 
PINN uses a deep neural network to approximate a function over physical space and introduces constraints such as PDEs and boundary conditions directly in the loss function to train the parameters in the neural network. 
The neural network is called ``physics informed'' because the parameters and boundary conditions are mostly related to physical quantities and the PDE is usually a physical law. 

In the present work, a PINN is used to approximate the specific intensity represented at given frequency points by vector $I_\nu=f(r,\varphi;w)$, where $w$ represents the trainable parameters in the neural network. 
To train the neural network, three sets of data points are sampled in the physical space $(r,\varphi)$: 1) the collocation points, where the PDE is enforced, are randomly sampled in the physical space from uniform distributions $r_{i,p} \in U(r_{\rm min},r_{\rm max})$ and $\varphi_{i,p} \in U(0,\pi)$; 2) the lower boundary points where $r_{j,l}=r_{\rm min}$ and $\varphi_{j,l}\in U(0,\pi/2)$; 3) the upper boundary points where $r_{k,u}=r_{\rm max}$, $\varphi_{k,u}\in U(\pi/2,\pi)$. 
The PDE collocation points are used in the left-hand side of Equation~(\ref{eq:opticrt}) and Equation~(\ref{eq:gammart}) to calculate the residual $R_{i,p}$, which is used in the loss function. 
The upper and lower boundary points are directly used to calculate the predicted specific intensities: $I_{k,u}=f(r_{\rm max},\varphi_{k,u},w)$, $I_{j,l}=f(r_{\rm min},\varphi_{j,l},w)$ which are then used to calculate the residuals with respect to the pre-defined boundary conditions $R_{j,l}$, $R_{k,u}$. The loss function is 

\begin{equation}\label{eq:loss}
    L=w_p \sum_{i,\nu}R_{i,p}^2+w_l \sum_{j,\nu}R_{j,l}^2 +w_u \sum_{k,\nu}R_{k,u}^2 \,\, , 
\end{equation}
where $w_p$, $w_l$, $w_u$ are weight parameters, which should be specified before training and the summation is over the data points labeled $i$ and the sampled frequency points labeled $\nu$ (i.e., spectral sampling pixels). 
The gradient of the loss function over the trainable parameters $\frac{\partial L}{\partial w}$ is calculated by reverse-mode automatic differentiation using the chain rule. 
Knowing the gradient, the trainable parameters can be adjusted with small steps in order to reduce the loss function. 
In practice, accelerated gradient-based algorithms (i.e., RMSprop \citep{Hinton2012RMSProp}, Adam \citep{Kigma2014Adam}) are used to increase the training efficiency and avoid local minima of the loss function. 
When the neural network setup is appropriate for the problem, the loss function will be close to zero after several iterations and the neural network solution will be close to the true specific intensity solution. 

PINN training is implemented in pytorch \citep{Paszke2019pytorch}. 
Section \ref{subsec:gammanet} introduces our solution for the gamma-ray radiative transfer in SN~Ia atmospheres. 
Section \ref{subsec:optical} gives the structure of the optical network, which is the major neural network for solving the optical radiative transfer problem, and reports example synthetic spectra. 

The spectra obtained from PINN are compared to those from the Monte-Carlo radiative transfer code TARDIS using the same ejecta structure and the spectra from the formal solution calculated by the method in Appendix \ref{sec:formal}. 
Section \ref{subsec:temperature} introduces an approach to accelerate the calculation of plasma temperature using Equation~(\ref{eq-PLTA-electron-scattering}). 
Section \ref{subsec:training} discusses the computation time and the training schedule of the PINN. 

\begin{figure}
    \includegraphics[width=0.47\textwidth]{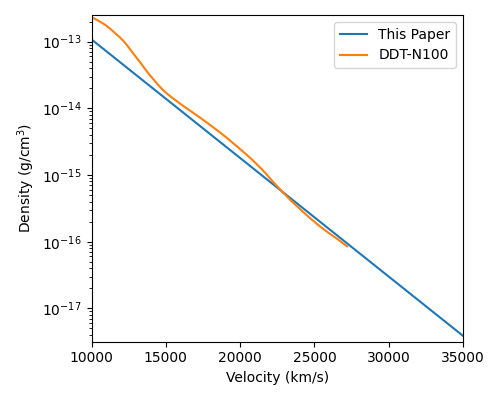}
    \includegraphics[width=0.47\textwidth]{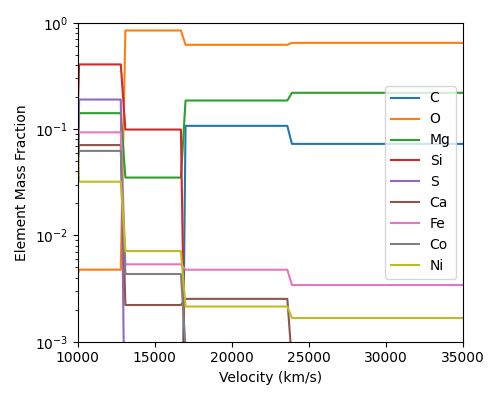}
    \caption{Upper Panel: The model SN~2011fe density  profile at 12.35 days after explosion used in the TARDIS and PINN calculations (blue line) and for comparison the density profile of model DDT-N100 \citep{Ropke2012N100} (orange line). 
    Lower Panel: The model SN~2011fe element mass fraction  at 12.35 days after explosion used in the TARDIS and PINN calculations.}\label{fig:SNexample}
\end{figure}

Our synthetic spectra are compared to the observed spectrum of SN~2011fe at 12.35 days after explosion. 
This observed spectrum \citep{Pereira2013FeSpec} was obtained by Double Spectrograph (DBSP) mounted on Palomar 200-inch (P200) Telescope. 
The observed spectrum was also used to derive the supernova ejecta structure via the method discussed in \cite{Xingzhuo2020AIAI}. 
The derived supernova ejecta density obeys
%
%
\begin{equation}
    \rho=3.87852\times 10^{-14}\times 0.689^{\frac{v-12500\,\mathrm{km/s}}{1000\,\mathrm{km/s}}}\,\mathrm{g/cm^{3}} \,\, , 
\end{equation}
where $v$ is the radial velocity in the ejecta. 
Figure \ref{fig:SNexample} shows the derived density and element abundance of the SN~Ia ejecta structure. 
We use only 9 elements, which are C, O, Mg, Si, S, Ca, Fe, Co, Ni, in this calculation for simplicity, and the highest ionization for these elements is limited to 3. 
Also in Figure \ref{fig:SNexample}, we compare our density profile with the density profile of model DDT-N100 \citep{Ropke2012N100}, which is also a SN~Ia ejecta model used to fit SN~2011fe spectra. 
Both the calculations in TARDIS and PINN set the lower and upper boundary to be at, respectively, the velocity coordinates $10151.4\, \mathrm{km/s}$ and $35675.3\, \mathrm{km/s}$. 

\subsection{The Gamma-Ray Network}\label{subsec:gammanet}

The gamma-ray network calculates the gamma-ray specific intensity $I_{\gamma}(r,\varphi)$ in the supernova atmosphere using Equation~(\ref{eq:gammart}) as the PDE. 
As a simplification, we assume the gamma-ray photon energies from the ${\rm ^{56}Ni}\rightarrow{\rm ^{56}Co}\rightarrow{\rm ^{56}Fe}$ decay chain are all 1 MeV to form the $j_{\rm r}$ term in Equation~(\ref{eq:gammart}) and the input gamma-ray photon from the boundaries are also 1 MeV. 
We neglect the kinetic energy and gamma-ray energy of the positron released in the ${\rm ^{56}Co}$ decay.
As discussed above in \ref{subsec:gamma}, we limit the allowed gamma-ray photon energies to discrete energy bins in the gamma-ray network as a simplification. 
There are 7 bins:  [1,0.407,0.243,0.171,0.131,0.106,0.088] MeV. 
Similar to the optical network, the gamma-ray network is trained on PDE collocation points and upper boundary points and lower boundary points. 
The upper boundary condition is no photons entering the SN~Ia atmosphere ($I_{\gamma}(r=r_{\rm max},\varphi\in [\pi/2,\pi])=0$).
The lower boundary condition is photons entering the supernova atmosphere from the lower boundary are in the highest energy bin ($I_{\gamma=1\mathrm{MeV}}(r=r_{\rm min},\varphi\in [0,\pi/2])=I_0$, $I_{\gamma<\mathrm{1\,MeV}}(r=r_{\rm min},\varphi\in [0,\pi/2])=0$). 
Two values of $I_0$ are used as specified below.

The best treatment of gamma-ray radiative transfer would be to use the time-dependent radiative transfer equation and set the inner boundary to be the supernova center. 
However, this treatment requires an extra input dimension on the neural network, which could considerably increase the training time. 
Moreover, the source term $j_{\rm r}$ and the intensity $I_{\gamma}$ would change several orders of magnitude throughout the whole supernova structure, and the neural network does not give good performance over multiple orders of magnitude. 
Therefore, the gamma-ray radiative transfer calculation is limited to supernova upper atmosphere and the effect of the gamma-rays from the supernova center is approximated with the lower boundary condition. 

We calculate two PINN models with different boundary conditions. 
In the first model, the inflow gamma-ray intensity is zero for both the upper and lower boundaries. 
In the second model, the inflow gamma-ray intensity from the lower boundary (i.e., $I_0$) is $10^{16}\,\mathrm{cm^{-2}s^{-1}sr^{-1}}$ with all gamma-ray energies set to 1\,MeV and no inflow of gamma-rays from the outer boundary. 
The neural network is a 10-layered fully-connected neural network, the number of neurons is $[2,128,128,128,128,128,128,128,256,256,7]$, and the activation function is the hyperbolic tangent (i.e., $\mathrm{tanh}$) for all the layers except the input and the output layers.
The input and output layers use a linear activation function. 

Note the original loss function in Equation~(\ref{eq:loss}) is not written with residuals in physically consistent units which makes assigning weights difficult.
Therefore, we write the residuals in terms of natural units as follows: 
\begin{equation}
    R_{i,p,{\rm new}}=\frac{R_{i,p}}{{\rm Mean}(k_{\rm C}+k_{\rm p})I_{\max}} \,\, , 
\end{equation}
\begin{equation}
    R_{j,l,{\rm new}}=\frac{R_{j,l}}{I_{\max}} \,\, , 
\end{equation}
\begin{equation}
    R_{k,u,{\rm new}}=\frac{R_{k,u}}{I_{\max}} \,\, , 
\end{equation}
where $I_{\max}$ is the maximum gamma-ray intensity and ${\rm Mean}(k_{\rm C}+k_{\rm p})$ is the mean opacity over all the PDE collocation points and all the gamma-ray energy bins. 
We set $I_{\rm max}$ using the equation
\begin{equation}
    I_{\max}=I_{l}+\int_{r_{\min}}^{r_{\rm max}} j_{\rm r} dr \,\, , 
\end{equation}
where $I_{l}$ is the lower boundary inflow intensity at 1 MeV, the $I_{l}$ value is zero for the first model and $10^{16}\,\mathrm{cm^{-2}s^{-1}sr^{-1}}$ for the second model, and the integral is over the source term in the supernova atmosphere. 
In the loss function Equation~(\ref{eq:loss}), $R_{i,p}$, $R_{j,l}$, $R_{k,u}$ are replaced by $R_{i,p,{\rm new}}$, $R_{j,l,{\rm new}}$, $R_{k,u,{\rm new}}$, respectively. 
Using this modification of the loss function, the order of magnitude of the three residual terms will not change drastically with the change of supernova ejecta model or the boundary conditions. 
Thus, the modification helps to balance the importance of PDE and boundary conditions when training the PINN. 
We found the PINN results are stable when the weight parameters in Equation~(\ref{eq:loss}) are $w_{p}=1,\ w_{l}=3000,\ w_{u}=3000$, respectively. 

\begin{figure}
    \plottwo{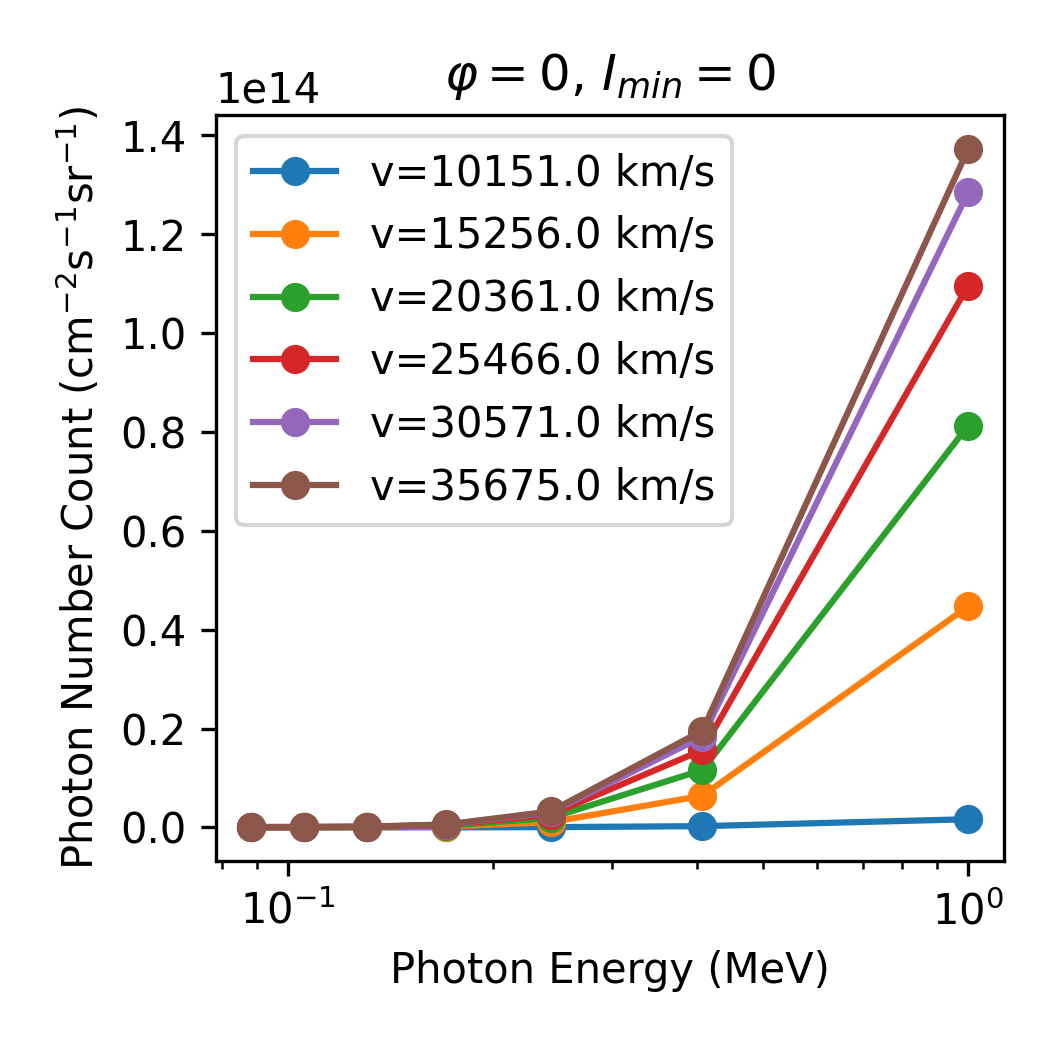}{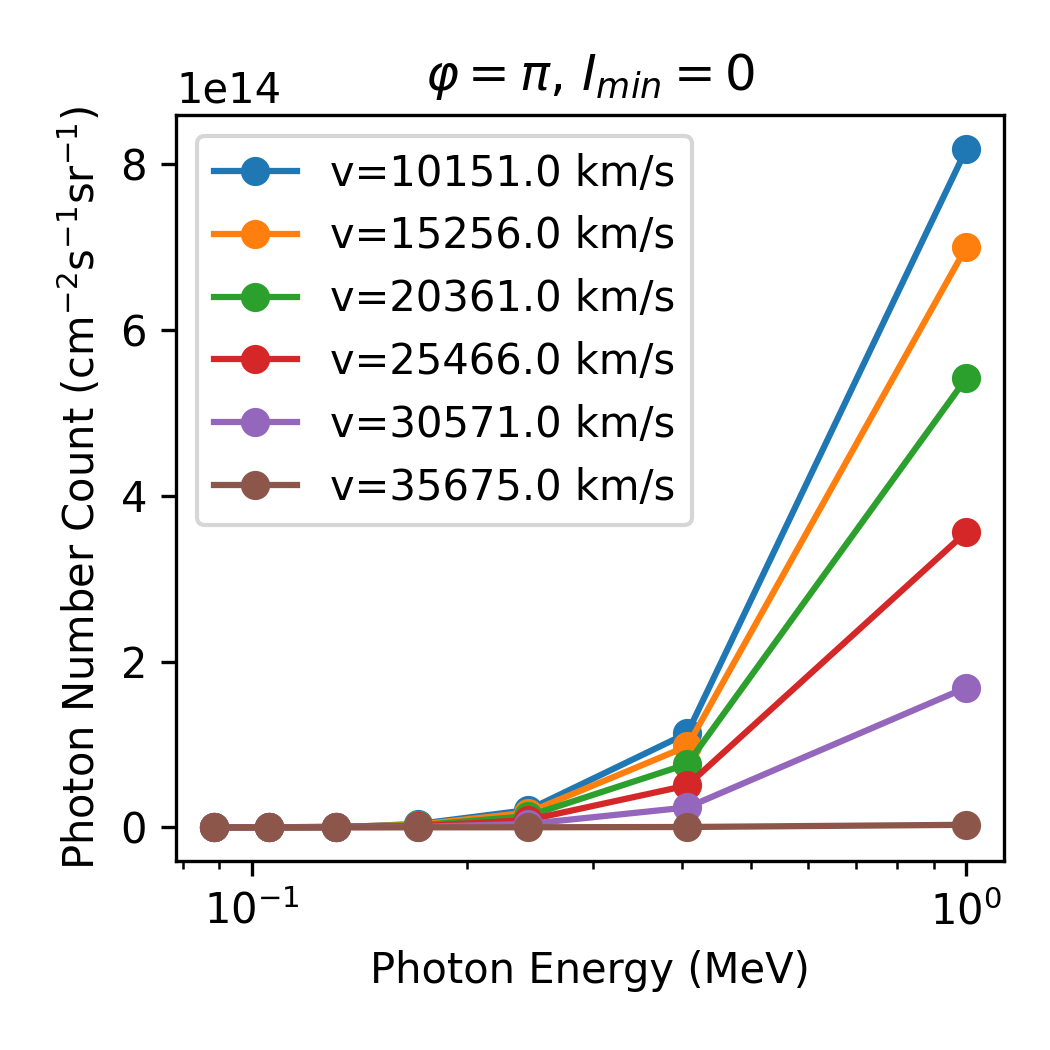}
    \plottwo{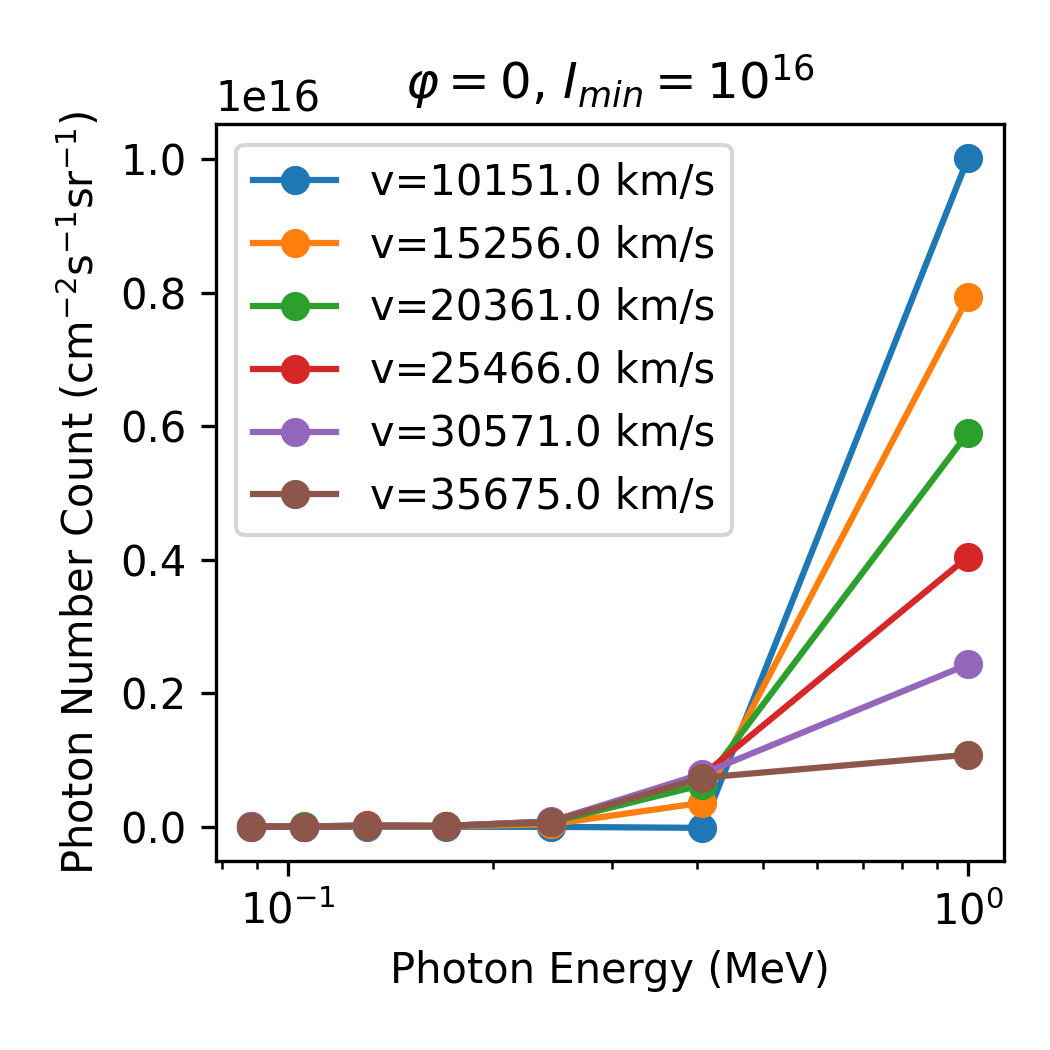}{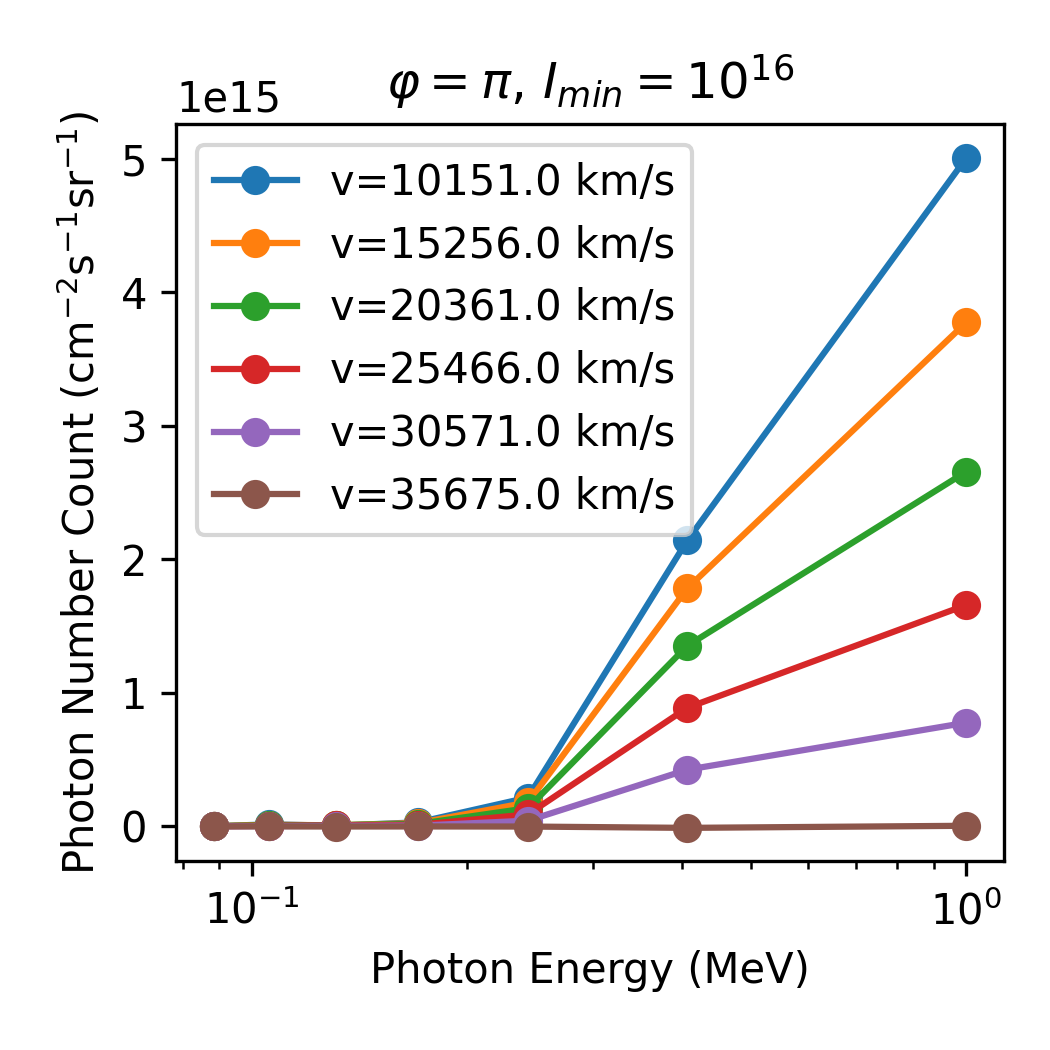}
    \caption{The gamma-ray specific intensity as a function of gamma-ray photon energy at representative velocity points and viewing angles in the two supernova ejecta models used to investigate gamma-ray radiative transfer in the ejecta. 
    The vertical axis unit is shown at the left side of each panel. 
    The upper panels show the results from the first model and the lower panels show the results from the second model. 
    The left panels show the results at viewing angle $\varphi=0$ and the right panels show the results at viewing angle $\varphi=\pi$. 
    }\label{fig:gammainten}
\end{figure}

Figure \ref{fig:gammainten} shows the results of the two models for gamma-ray radiative transfer. 
Note that both models observe the lower and upper boundary conditions accurately. 
In the first model, we note that the intensity at the viewing angle $\varphi=0$ and $1\,$MeV energy bin increases with the increase of radial velocity, due to the ${\rm ^{56}Ni}$ and ${\rm ^{56}Co}$ in the supernova atmosphere shown in the lower panel of Figure \ref{fig:SNexample}. 
In the second model, we note the intensity is much larger than that of the first model, which means the input energy from the lower boundary dominates over the radioactive energy in the supernova atmosphere. 
We note that there is relatively little scattering of gamma-ray photons to energy bins belov $1\,$Mev in all cases. 
The most such scattering occurs for the second model at the viewing angle $\varphi=\pi$. 
This is to be expected since inward moving gamma-ray photons will be mostly scattered gamma-ray photons when the gamma-ray intensity is dominated by a central source. 

\begin{figure}
    \plotone{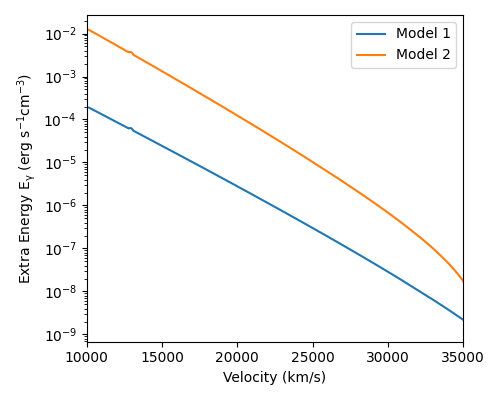}
    \caption{The gamma-ray energy $E_{\gamma}$ calculated from the two supernova ejecta models used to investigate gamma-ray radiative transfer in the ejecta. 
    The model names are labeled in legend. }\label{fig:extengexample}
\end{figure}

Figure \ref{fig:extengexample} shows the gamma-ray energy deposition $E_\gamma$ as a function of radial velocity for the two models. 
Note the gamma-ray radiative transfer equation (Eq.~(\ref{eq:gammart}) is linear since the gamma-ray sources and opacities are fixed inside the atmosphere. 
Therefore, the general gamma-ray intensity (or energy deposition) for our supernova ejecta structure with a general inner boundary condition can be written as a linear combination of our two PINN-calculated gamma-ray intensity solutions (or energy deposition solutions). 
To be explicit for the general gamma-ray energy deposition, one has
\begin{eqnarray}
    E_{\gamma}(I_{l},r)&=&E_\gamma (I_{l,1},r)+\frac{I_{l}-I_{l,1}}{I_{l,2}-I_{l,1}}\left[E_\gamma(I_{l,2},r)-E_\gamma(I_{l,1},r) \right] \cr
    &=&E_\gamma (I_{l,1},r)+\frac{I_{l}}{I_{l,2}}\left[E_\gamma(I_{l,2},r)-E_\gamma(I_{l,1},r) \right] \,\, ,
\end{eqnarray}
where $I_{l,1}=0$ is the lower boundary condition of the first model, $I_{l,2}=10^{16}\,\mathrm{cm^{-2}s^{-1}sr^{-1}}$ is the lower boundary condition of the second model, and $I_{l}$ is the target lower boundary condition. 
Using this relation, the gamma-ray energy deposition for our supernova ejecta structure with different inner boundary conditions can be generated without extra PINN calculations. 

\subsection{The Optical Network}\label{subsec:optical}

The optical neural network calculates the optical specific intensity $I_{r,\varphi}$ using Equation~(\ref{eq:opticrt}) as PDE. 
The output of the PINN is the specific intensity sampled in a frequency grid between $10^{14.4}\,\mathrm{Hz}$ ($11,935\,\mathrm{\AA}$) and $10^{15}\,\mathrm{Hz}$ ($3000\, \mathrm{\AA}$) with 2048 pixels uniformly sampled in the logarithmic space. 
The frequency upper limit is set as $10^{15}\,\mathrm{Hz}$ for two simplification reasons. 
First, several strong Fe-group element spectral lines, which could lead to order-of-magnitude problems in the training of PINN, lie above this frequency. 
Second, bound-free opacity is more significant above this frequency, and we have not included bound-free opacity in general as simplification. 
In fact, the specific intensity is suppressed by the high-opacity spectral lines and the bound-free opacity above $10^{15}\,\mathrm{Hz}$. 
Therefore, removing the specific intensity calculation above this frequency will probably not lead to significant error in a direct sense. 
However, the thermal state of the ejecta can only be crudely approximated without the high frequency region, bound-free opacity, and NLTE effects. 

Considering the output of the neural network is a vector with the length of 2048, while other PINN applications have typically one or a few dimensions as output \citep[e.g.,][]{MMPINNRTE2021}, we prepared a large 14-layered neural network structure (hereafter N14). 
The number of neurons in each layer is, respectively, [2, 256, 256, 256, 256, 256, 512, 512, 512, 512, 512, 2048, 2048, 2048, 2048] and the activation function is the hyperbolic tangent (i.e., $\mathrm{tanh}$) for all the layers except the input and the output layers: the input and output layers use a linear activation function. 
We also prepared a smaller 6-layered neural network (hereafter N6) for comparison: the number of neurons in each layer is [2, 512, 2048, 2048, 2048, 2048]. 

The upper boundary condition is that no radiation flow into the material ($I(r_{\rm max},\varphi\in [\pi/2,\pi])=0$) and the lower boundary condition is that the input radiation flow is an isotropic blackbody spectrum with a pre-defined boundary temperature $T_{\rm Bo}$: 
\begin{equation}\label{eq:opticlower}
    I_{r_{\rm min},\varphi\in [0,\pi/2]}=\frac{2h\nu^3}{c^2\left(e^{\frac{h\nu}{k_{\rm B}T_{\rm Bo}}}-1\right)} \,\, . 
\end{equation}
We found the synthetic spectra are close to the observed spectra when $T_{\rm Bo}=11500 \mathrm{K}$, and thus we adopted this value for all our calculations. 
As a simplification, we used $I_{r_{\rm min}}$ as a rest frame specific intensity though. 
Formally it should be a comoving frame specific intensity. 
The distinction between the two quantities is small. 

Similar to the gamma-ray network, residuals in the loss function Equation~(\ref{eq:loss}) are written in natural scaled versions: 
\begin{equation}
    R_{i,p,{\rm new}}=\frac{R_{i,p}}{I_{\rm max}{\rm Mean}(k_{\rm e}+k_{\rm bb})} \,\, , 
\end{equation}
\begin{equation}
    R_{j,l,{\rm new}}=\frac{R_{j,l}}{I_{\max}} \,\, , 
\end{equation}
\begin{equation}
    R_{k,u,{\rm new}}=\frac{R_{k,u}}{I_{\max}} \,\, , 
\end{equation}
where $I_{\max}$ is the maximum pixel value of the lower boundary condition Equation~(\ref{eq:opticlower}) and the mean is over all the frequency pixels and PDE collocation points. 
The weight parameters used in Equation~(\ref{eq:loss}) are $w_p=1$, $w_l=3000$, and $w_u=3000$. 
These are the same as for the gamma-ray radiative transfer calculation and were found comparably good for the optical spectrum calculations. 

For our example synthetic spectrum calculation, we use the gamma-ray energy $E_{\gamma}$ calculated in Section \ref{subsec:gammanet} and assume the gamma-ray intensity lower boundary condition is $I_{0}=10^{16}\, \mathrm{cm^{-2}s^{-1}sr^{-1}}$ and the plasma temperature is calculated using the temperature network, which is described in \S~\ref{subsec:temperature}. 

\begin{figure*}[htb!]
    \plottwo{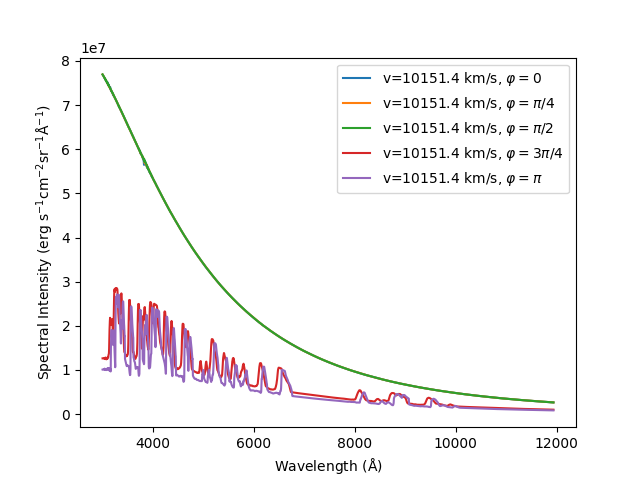}{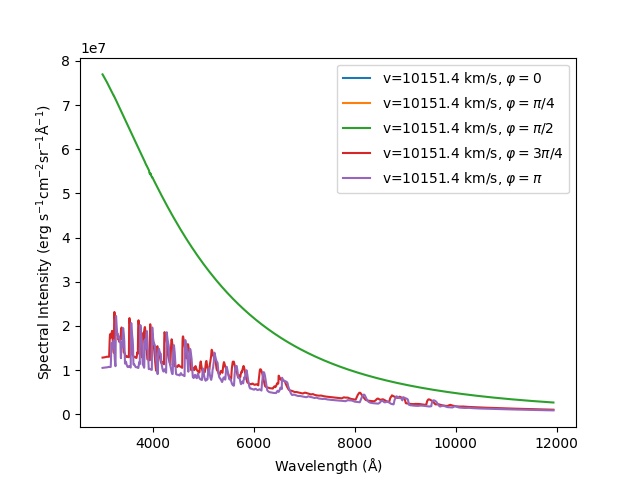}
    \plottwo{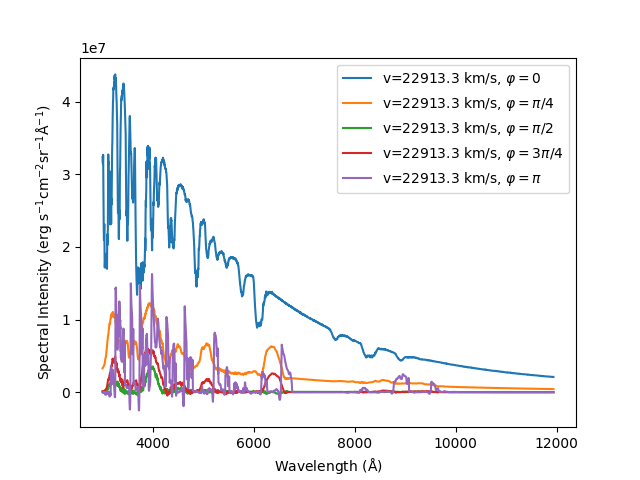}{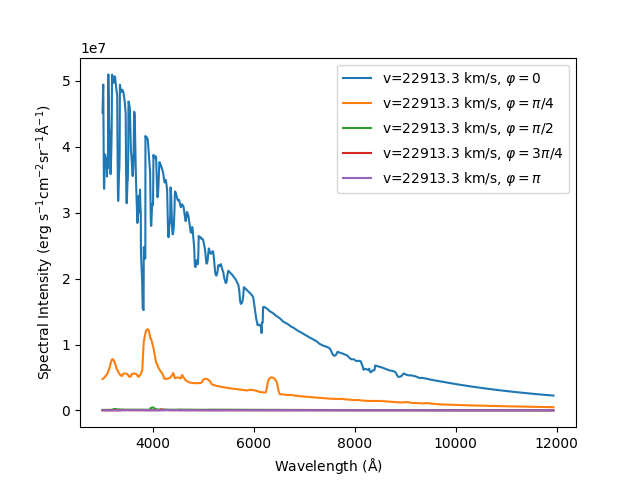}
    \plottwo{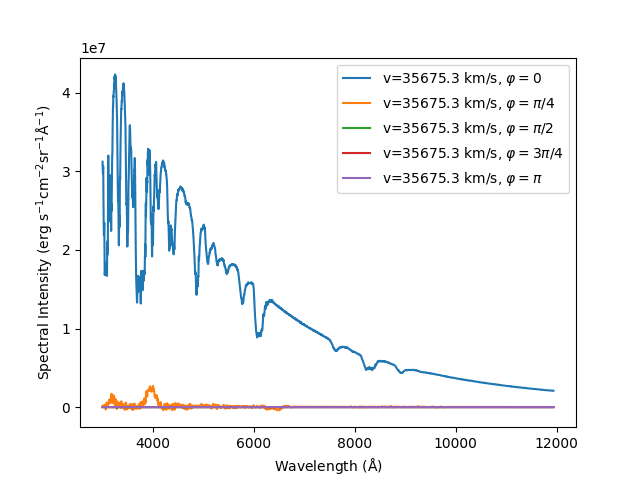}{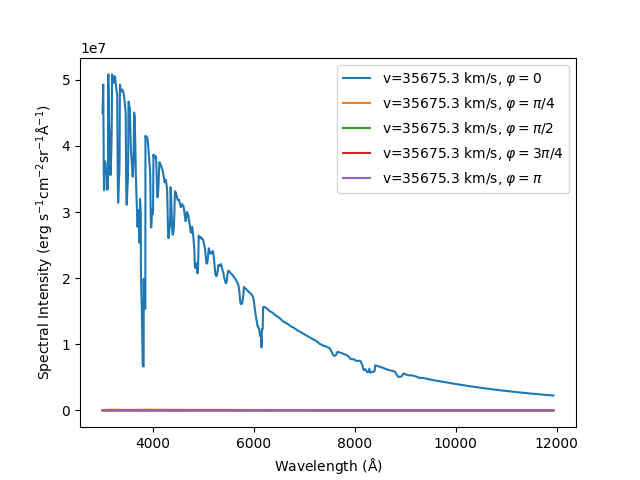}
    \caption{The rest frame specific intensity as a function of rest frame wavelength at representative radial velocities and viewing angles. 
    Note specific intensity is in the wavelength representation rather then in the frequency representation which we use in the text. 
    The left panels show specific intensities from PINN calculations using the N14 neural network structure and the right panels show the corresponding specific intensities from the formal solution of the PDE. 
    The coordinates and viewing angles are shown in legends.
    In the upper panels, the specific intensity curves for $\varphi \leq \pi/2$ are just the inner boundary specific intensity, and so all overlap and give a net green color.
    In the middle right panel, the specific intensity curves for $\varphi \geq \pi/2$ are all nearly zero, and so overlap and give a net purple color. 
    In the lower left panel (lower right panel), the specific intensity curves for $\varphi \geq \pi/2$ ($\varphi \geq \pi/4$) are all nearly zero, and so overlap and give a net purple color.
    }\label{fig:opticinten}
\end{figure*}

Figure \ref{fig:opticinten} shows the specific intensities as functions of rest frame wavelength sampled at representative radial velocities and viewing angles calculated by the N14 neural network (i.e., the N14 PINN) and the formal solution (see Appendix \ref{sec:formal}). 
Note that at the lower boundary ($v=10151.4\,\mathrm{km/s},\ \varphi<\pi/2$) and the upper boundary ($v=35675.3\,\mathrm{km/s},\ \varphi>\pi/2$), the PINN solutions satisfy the boundary condition virtually exactly as should be the case. 
Also note that there were no specific intensity values less than zero which should be the case. 
Blueshifted absorption lines can be observed in the $\varphi=0$ spectra and emission lines of varying shift can be seen in the spectra with $\varphi>0$. 

To explicate the absorption lines in the specific intensity spectra, note the following equation for rest frame wavelength derived from the Doppler shift formula Equation~(\ref{eq:Dopplershift}):
\begin{equation}\label{eq:Dopplershift_2}
    \lambda=\bar\lambda_{\rm line}\gamma[1-\mathrm{cos}(\varphi)\beta] \,\, , 
\end{equation}
where $\bar\lambda_{\rm line}$ is the comoving frame line wavelength. 
Consider a beam of rest frame wavelength $\lambda$ (which is invariant of course as the beam propagates) from the inner boundary at a point~A that reaches a point~B in the atmosphere where we evaluate the rest frame specific intensity at that wavelength $\lambda$. 
Given that the beam rest frame wavelength $\lambda$ is sufficiently (and not too much) blueward of a line wavelength $\lambda_{\rm line}$, it will redshift in the comoving frame to $\bar\lambda_{\rm line}$ and be partially absorbed at a point~C along the beam path before reaching point~B. 
Point~C has radial velocity coordinate $\beta=v/c$ and viewing angle $\varphi$. 
Note beams from the inner boundary that reach point~B always have $\varphi\leq\pi/2$, and so $\mathrm{cos}(\varphi)\beta\geq0$ and $\lambda\leq\lambda_{\rm line}$, and there is a blueshift of the absorption relative to $\lambda_{\rm line}$ in the rest frame. 
(Note also $\gamma\approx1$ for virtually all cases for supernovae.)
Since there is a range of radial velocity coordinate $\beta=v/c$ values and viewing angle $\varphi$ values along the beam path from point~A to point~B where beams redshift into the line in the comoving frame to $\bar\lambda_{\rm line}$, the absorption in the specific intensity spectrum is a broad absorption line all blueward of rest frame wavelength $\lambda_{\rm line}$.
There is also line emission along the beam path, but the absorption usually dominates for beams originating on the inner boundary. 
A detailed explication of blueshifted absorption features and supernova spectral line formation in general is given by \citet[][173--194]{jefferybranch1990}. 

\begin{figure*}[htb!]
    \plottwo{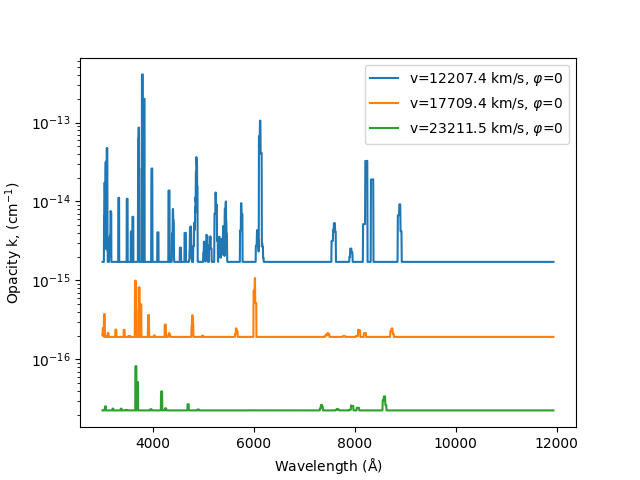}{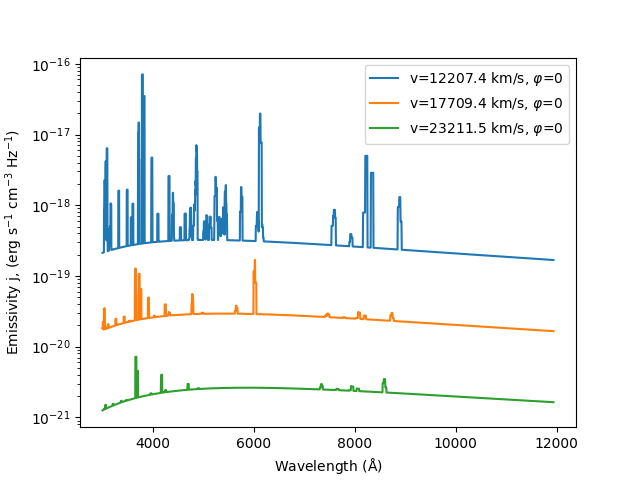}
    \caption{Left panel is the rest frame opacity $k$ at various velocity coordinates. 
    Right panel is the rest frame emissivity $j$ at various velocity coordinates. 
    The wavelength is rest frame wavelength.}\label{fig:jkexample}
\end{figure*}

To explicate the emission lines in the specific intensity spectra, consider the same setup as in the last paragraph, except with the beam path starting from a point~A on the outer boundary. 
(The setup is, in fact, illustrated in Figure \ref{fig:illuFormal} in Appendix \ref{sec:formal}.)
In this case, there is only net emission of radiation at comoving frame wavelength $\bar\lambda_{\rm line}$ along the beam path from point~A to point~B. 
For our explication, it is best to think of the rest frame wavelength $\lambda$ at point~B as a variable depending on $\mathrm{cos}(\varphi)\beta$ evaluated at a general point of emission along the beam path. 
There are two cases depending on $\varphi_{\rm B}$, the viewing angle at point~B: first case,
$\varphi_{\rm B}<\pi/2$ and second case, $\varphi_{\rm B}\geq\pi/2$. 
In the first case, the emission at a general point gives redshifted (unshifted, blueshifted) specific intensity relative to $\lambda_{\rm line}$ at point~B for $\varphi>\pi/2$ ($\varphi=0$, $\varphi<\pi/2$). 
The emission line in the spectrum will be broad in wavelength and will straddle the line wavelength $\lambda_{\rm line}$. 
Since the beam path is closest to the inner boundary for $\varphi=0$, the emission will usually be strongest at this point, and so the emisison line will usually peak at $\lambda_{\rm line}$ in the spectrum. 
Since the emission heading toward point~B along the beam path abruptly stops at point~B, the blue end of the emission feature in the spectrum will usually show a steep rise with wavelength. 
The red edge will be a less steep fall since the redshifted emission will usually have a gentler decline as the emission point recedes through less strongly emitting matter as it recedes to the outer boundary of the atmosphere. 
Now in the second case (where $\varphi_{\rm B}\geq\pi/2$), the emission along the beam path is usually strongest right at point~B itself since all other emission points are at greater radii than point~B. 
So the blue edge of the emission feature will have a steep rise to the emission peak usually.
The emission peak in the spectrum will be at $\lambda_{\rm line}$ for $\varphi_{\rm B}=0$ and will shift redward as $\varphi_{\rm B}$ increases usually. 

The absorption and emission lines seen in the spectra in Figure \ref{fig:opticinten} are consistent with explications given in the last two paragraphs. 
However, when comparing the PINN spectra to the formal solution spectra, we notice significant differences at coordinates ($v=22913.3\,\mathrm{km/s},\ \varphi\geq\pi/2$) and ($v=35675.3\,\mathrm{km/s},\ \varphi=\pi/4$). 
The PINN spectrum emission lines are much larger than the formal solution emission lines (which in fact are close to zero). 
The same discrepancy also occurs when the PINN spectra are calculated by the N6 neural network structure. 

To investigate the cause of the discrepancies between the PINN and formal solution specific intensity spectra, we have plotted in Figure \ref{fig:jkexample} several example rest frame opacity and emissivity terms at different coordinates as functions of rest frame wavelength. 
We note that the opacity and emissivity due to the bound-bound transitions (which are seen as sharp spikes) can be up to $\sim 1000$ times the corresponding electron scattering terms for $v=12207.4\ \mathrm{km/s}$ and up to $\sim 10$ times the corresponding electron scattering terms
for $v=23211.5\ \mathrm{km/s}$. 
We surmise the discrepancy between the PINN solution and the formal solution for the specific intensity spectra is due to the large order-of-magnitude variation in the opacity and emissivity terms in the PDE loss function that goes beyond the dynamic range of neural networks. 

\begin{figure*}
    \plotone{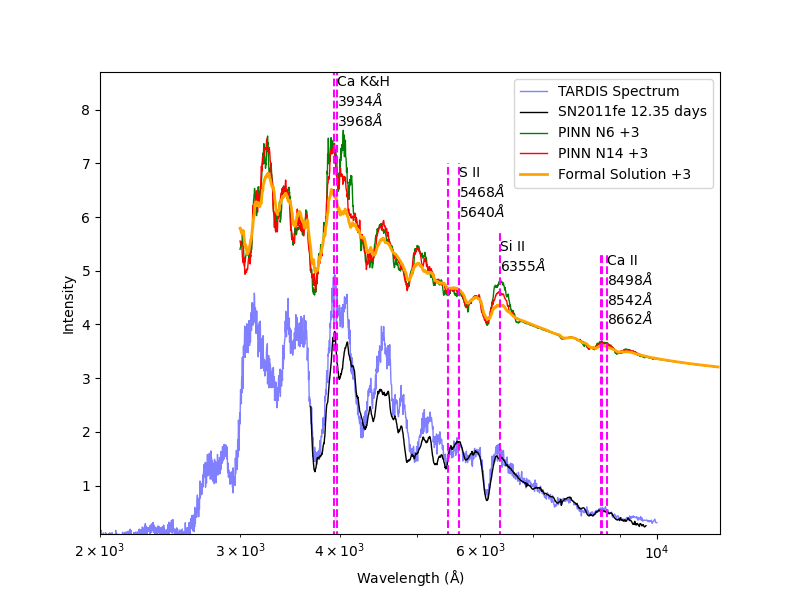}
    \caption{In the wavelength representation, the spectrum of SN~2011fe at 12.35 days after explosion (black line), the TARDIS synthetic spectrum (blue line), the PINN spectra from N14 and N6 neural network structures (red line and green line), and the formal solution spectrum (orange line). 
    The intensity is in arbitrary units and wavelength is on a logarithmic scale.
    The PINN spectra and formal solution spectrum are moved upward by 3 units (as indicated in the legend) for clarity. 
    Several spectral lines are marked with magenta dashed lines.}\label{fig:examplespec}
\end{figure*}

Using the specific intensity calculated by PINN or the formal solution, a synthetic spectrum that can be compared to observations is calculated by the integral
\begin{equation}
    {\rm Spec}(\nu)=\int_{0}^{\pi/2}d\varphi\ 2\pi r_{\rm max}^2 I(r_{\rm max},\varphi,\nu)\mathrm{sin}(\varphi)\mathrm{cos}(\varphi) 
\end{equation}
\citep[e.g.,][p.~11--12]{Mihalas1978Book}. 
%
%
In the wavelength representation (not the frequency representation), Figure \ref{fig:examplespec} shows two PINN synthetic spectra from the N14 and N6 neural network structures, the corresponding formal solution synthetic spectrum, the TARDIS synthetic spectrum (fitted to the observations using the method of \cite{Xingzhuo2020AIAI}), and the observed spectrum for SN~2011fe at 12.35 days after explosion. 
Because the method in \cite{Xingzhuo2020AIAI} is specifically designed for TARDIS, as well as the supernova ejecta structure used in this paper, the TARDIS synthetic spectrum fits observed spectrum with reasonable accuracy. 
However, the PINN and formal solution spectra (which are obtained without the optimized fitting of \cite{Xingzhuo2020AIAI}) do not fit to the same level of accuracy and cannot be expected to do so. 

The test of the PINN synthetic spectrum calculation is the comparison to the formula solution synthetic spectrum calculation which is based on exactly the same atmosphere structure and thermal state and is calculated with guaranteed numerical accuracy. 
First, we note the PINN spectra and the formal solution spectrum are qualitatively alike. 
In particular, they both exhibit typical P-Cygni line profiles as are also seen in the observed and TARDIS spectra. 
P-Cygni line profiles are characteristic of expanding atmospheres and have a broad emission feature centered around the spectral line center and a blueshifted absorption feature. 
Line blending can distort P-Cygni line behavior to unrecognizablity. 
In the spectra in Figure \ref{fig:examplespec}, P-Cygni line profiles are recognizable for several conspicuous spectral lines: Ca K\&H $3934\, \mathrm{\AA}$, $3968,\mathrm{\AA}$, Si II $6355\,\mathrm{\AA}$, S II $5468\,\mathrm{\AA}$ (multiplet average), S II $5640\,\mathrm{\AA}$ (rough average of several lines); $\mathrm{Ca\ II\ } 8498\,\mathrm{\AA}, 8542\,\mathrm{\AA}, 8662\,\mathrm{\AA}$. 
Qualitatively, the agreement between the PINN spectra and the formal solution spectrum is moderate. 
Overall, the PINN spectra show stronger emission features. 
This is to be expected given that the PINN line emission specific intensities in Figure \ref{fig:opticinten} were generally too strong in the PINN case. 
The PINN spectra also show noise which is to be expected for PINN calculations. 
We note that neither the PINN nor the formal solution produce the ``W'' shaped S II feature around $5500\, \mathrm{\AA}$ seen in the observed spectrum. 
The TARDIS spectrum does produce this shape qualitatively and this is probably attributable to TARDIS's better thermal state calculation compared to ours.
TARDIS uses a dilute-blackbody approximation in the temperature calculation and the macroatom approximation in the source function calculation. 
Our thermal state calculation is simpler and is described in \S\S~\ref{sec:rte} and~\ref{subsec:temperature}.

\subsection{The Temperature Network}\label{subsec:temperature}

The plasma temperature calculated directly from Equation~(\ref{eq-PLTA-electron-scattering}) requires numerical integration. 
However, it is known from previous simulations \citep[e.g.,][]{Xingzhuo2020AIAI}, that the temperature profile in supernovae above the photosphere is a smooth function of radial velocity. 
Therefore, we use a simple neural network to interpolate the temperature profile during the PINN calculation in order to curtail the computational time in numerical integration. 
The neural network is a simple fully-connected neural network, the number of neurons per layer is [1,64,64,64,1], and the activation function is the SELU function \citep{Klambauer2017SELU}. 
During the training of the optical network, the radius values of the PDE collocation points are input into the temperature neural network, then the predicted temperature values are used to calculate the level populations and the opacity and emissivity in Equation~(\ref{eq:opticrt}). 
During the training of the temperature network, we randomly sample 200 radius values, then use the following loss function to train the neural network: 
\begin{equation}
    \begin{split}
    L_{T}=\sum_{m}\Big[
    4\sigma_{\rm SB}\sigma_{\rm T}N_{e} T(r_m)^4-E_{\gamma}(r_m)-  \\
    \sigma_{\rm T}N_{e}\int_{\nu_{\rm min}}^{\nu_{\rm max}}d\nu\int_{\Omega}d\Omega\, I(r_{m},\varphi) \Big]^2 \,\, , 
    \end{split}
\end{equation}
where $r_m$ is the radius from the random sample, $T(r_{m})$ is the neural network predicted temperature, and the integral is calculated using trapezoid integration over the 2048 frequency pixels and Monte-Carlo integration with 200 sampling points over the viewing angle $\varphi$. 
The loss function forces an approximate solution of Equation~(\ref{eq-PLTA-electron-scattering}) throughout the atmosphere. 
Note in the above equation we used $I(r_{m},\varphi)$ which is a rest frame specific intensity. 
Properly, we should use a comoving frame specific intensity. 
However, the distinction between the two quantities is small. 

\begin{figure}
    \plotone{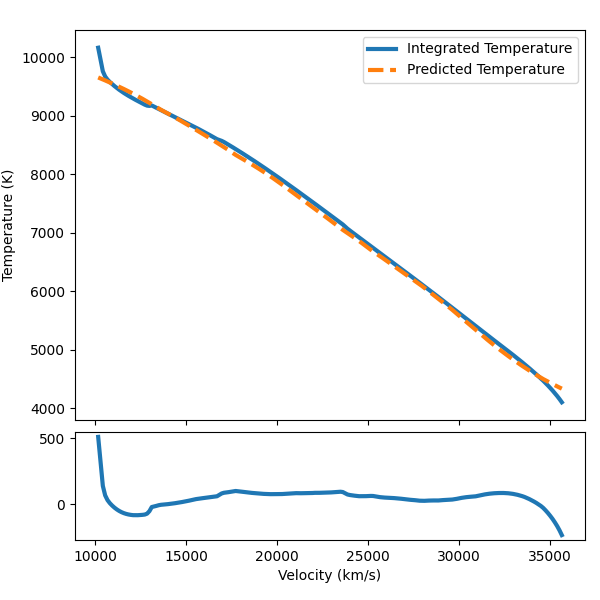}
    \caption{Upper panel: The predicted temperature from the temperature neural network (orange dashed line) and the temperature integrated from the specific intensity using Equation~(\ref{eq-PLTA-electron-scattering}) as functions of radial velocity. 
    Lower panel: The difference between two temperatures. }\label{fig:temperexamp}
\end{figure}

Figure \ref{fig:temperexamp} shows the neural network predicted temperature and the temperature calculated from the integration of the specific intensity $I$. 
There is reasonably good agreement between the two temperature curves as seen in the lower panel. 
At most radial velocities, the temperature difference is smaller than $50\,\mathrm{K}$, while the temperature difference near the lower and the upper boundary is about $300$--$500\,\mathrm{K}$. 
Although other methods (i.e., linear interpolation, cubic spline) may be as good as, or even better than, the neural network in approximating the temperature profile as a function of radius $T(r)$ from 200 sampling points, the neural network will be a better interpolation function in higher dimensional problems (e.g., the 3D radiative transfer problem). 
So we will continue to use the neural network as the temperature interpolation function for the upgrades in the future. 

\subsection{The Training Procedure}\label{subsec:training}

When training the gamma-ray neural network, we use the Adam algorithm \citep{Kigma2014Adam} to update the trainable parameters. 
In each epoch of training, the neural network reads a small batch of data with 2000 PDE collocation points, 2000 upper boundary points, and 2000 lower boundary points to update the trainable parameters. 
An epoch has 200 batches and there are 140 epochs. 
During the training, the learning rate changes from $1\times10^{-4}$ to $1\times 10^{-6}$. 
The first few epochs have larger learning rates to efficiently train the neural network to an approximate solution, and then the smaller learning rates in the following epochs increase the training precision. 
The total training time is about 36 minutes using one Nvidia-A100 GPU card. 

The optical neural network with the N14 structure is much more sophisticated and requires more computation time: 62 hours in total (see below). 
The computation time of different sub-steps of the training on a Nvidia-A100 GPU card are as follows: 

\begin{itemize}
    \item Calculating the $k_{\rm abs}$ and $j_{\rm em}$ values of 1500 PDE collocation points on 2048 frequency sampling pixels takes $965\,$ms. 
    \item Updating the optical neural network with the Adam algorithm using 1500 PDE collocation points, 1500 upper boundary points, and 1500 lower boundary points takes $71\,$ms. 
    \item Calculating the integral in Equation~(\ref{eq-PLTA-electron-scattering}) over 200 sampling points takes $863\,$ms. 
    \item Updating the temperature with the Adam algorithm using 200 sampling points 
    (see \S~\ref{subsec:temperature}) takes $4.5\,$ms. 
\end{itemize}

Note the calculation of the $k_{\rm abs}$ and $j_{\rm em}$ values and the integral in Equation~(\ref{eq-PLTA-electron-scattering}) take much longer times than the other two computation processes. 
In order to accelerate the training, we make two modifications. 
First, we repeat the updating of optical neural network 20 times on the same batch of data points. 
Second, because the temperature neural network is much simpler than the optical neural network, we set the learning rate to be 10 times of that of the optical neural network. 

As aforesaid for the optical network, the training batch size is 1500 for the PDE collocation points and for each of the upper and lower boundary points. 
An epoch has 400 batches and the total training procedure has 136 epochs. 
Similar to the gamma ray neural network, the learning rate of the optical neural network changes from 0.0001 to $1\times 10^{-6}$ during the training epochs. 
The total training time of the N14 neural network is 62 hours using one Nvidia-A100 GPU. 
The total training time of the N6 neural network is 60 hours using one Nvidia-A100 GPU, which means a simplified neural network structure did not significantly reduce the large amount of training time in the present work. 

\vspace{3ex}
\section{Conclusion}\label{sec:discuss}

We used PINNs to calculate an optical spectrum of Type Ia SNe SN~2011fe at 12.35 days after explosion. 
The specific intensity throughout the supernova atmosphere is roughly solved and the synthetic spectrum is in qualitative agreement with the observed spectrum and the formal solution spectrum, noting especially that the spectral line profiles caused by several important atomic transitions (e.g., $\mathrm{Si\ II\ }\ 6355\,\mathrm{\AA}$;  $\mathrm{Ca\ II\ } 8498\,\mathrm{\AA}, 8542\,\mathrm{\AA}, 8662\,\mathrm{\AA}$) are qualitatively reproduced. 

However, there are several challenges to the further exploration of the supernova explosion mechanism via the PINN-based method. 
First, the PINN-based method is inefficient at integration. 
The only integral calculation in the current PINN setup is equation~(\ref{eq-PLTA-electron-scattering}) for temperature which requires a significant amount of the computation time. 
%
%
%

Second, apart from the integration calculations, PINN calculation is slow. 
Despite the temperature neural network, the refined training strategy, and other tricks we have introduced, which have already accelerated the training procedure significantly, the computation cost of a full
simulation is about several GPU-days
In contrast, TARDIS typically uses several CPU hours to run a simulation. 

Third, the PINN spectrum is not quantitatively accurate as shown by comparison to the formal solution spectrum. 
We surmise that this is due to large order-of-magnitude variations in emissivity $j_{\mathrm{em}}$ and opacity $k_{\mathrm{abs}}$ (see \S~\ref{subsec:optical}). 
Using XPINN \citep{Hu2021xpinn}, which can separate the parameter space into different subdomains and connect the neural networks in different subdomains with extra boundary conditions, may alleviate this order-of-magnitude variation problem. 
However, we did not attempt this method in this paper because it can drastically increase the computational resources required. 

To summarize, using PINN in the forward modeling problem of supernova radiative transfer calculation faces multiple challenges in computational efficiency, and therefore in applying it to a large grid of supernova ejecta models. 
The challenges to PINN radiative transfer equally apply to the construction of a PINN inverse problem solver which encodes the supernova ejecta structure parameters into the input of the PINN and fits observed spectra. 
If the challenges are not overcome, an inverse problem solver will be too computationally demanding for use. 
The high dimensionality of the parameter space for an inverse problem solver adds to the challenges.

To summarize, innovative upgrades are necessary to significantly accelerate the PINN training process and radiative transfer calculation for either forward or inverse modeling. 
Those upgrades may include combining PINN with other methods:  e.g., Monte-Carlo or traditional numerical PDE methods. 

\section*{Acknowledgements}
    Portions of this research were conducted with the advanced computing resources provided by Texas A\&M High Performance Research Computing. 
    Ulisses Braga-Neto and Ming Zhong are supported by the National Science Foundation through NSF award CCF-2225507. 
    Lifan Wang acknowlege the NSF grant AST-1817099 for supports on this work.

\bibliography{RTPIref}{}
\bibliographystyle{aasjournal}

\appendix

\section{The formal solution of the radiative transfer equation}\label{sec:formal}

\begin{figure}
    \plotone{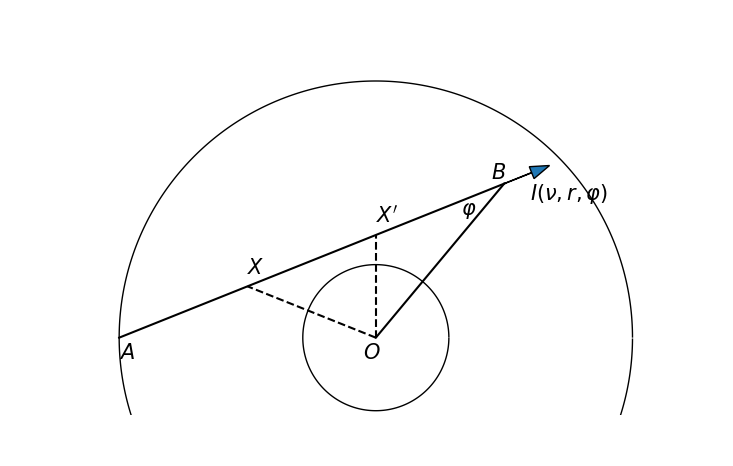}
    \caption{An illustration of the geometry of the atmosphere and the beam path used in our presentation of the formal solution. }\label{fig:illuFormal}
\end{figure}

In this appendix, we present the analytical formal solution of the specific intensity $I$ at a given coordinate $r$ and a viewing angle $\varphi$ for the radiative transfer equation (Eq.~(\ref{eq:opticrt})) written in terms of beam path coordinate $x$ where $x$ increases in the direction of radiation flow (i.e., the beam path direction). 
For our presentation, Figure \ref{fig:illuFormal} illustrates the geometry of the atmosphere and the beam path for the case that the beam path intersects the outer boundary of the atmosphere. 
Note the viewing angle $\varphi$ is the angle between outward radial direction and the beam path: we leave it unsubscripted for point $B$ and subscripted by $x$ and $x'$ for the corresponding points shown in Figure \ref{fig:illuFormal}. 
In terms, of $x$, the radiative transfer equation (neglecting time dependence) is
\begin{equation}\label{eq:Castor01_03}
    \frac{\partial I}{\partial x}
    -j_{\rm em}\left(\frac{\bar{\nu}}{\nu}\right)^{-2}
     +k_{\rm abs}\left(\frac{\bar{\nu}}{\nu}\right)I=0 \,\, ,
\end{equation}
where the frequency dependence is implicit (\citealt[e.g.,][eq.~(1--3)]{Castor1972RTE}; see also \citealt[][p. 31,33,495--496]{Mihalas1978Book}).
As a simplification, we define the opacity and emissivity in rest-frame as $K=k_{\mathrm{abs}}\left(\frac{\bar{\nu}}{\nu}\right)$ and $J=j_{\mathrm{em}}\left(\frac{\bar{\nu}}{\nu}\right)^{-2}$, and note that they both depend on the  viewing angle $\varphi$ via the $\bar{\nu}/{\nu}$ factor as seen from equation~(2) in \S~2. 
The formal solution follows straightforwardly using the integrating factor $e^{\int_A^{x} K(r_{x'},\varphi_{x'})\,dx'}$:
\begin{equation}
    I(\nu,r,\varphi)=I_{\rm BC}\ e^{-\int_A^{B} K(r_x,\varphi_x)\,dx}+\int_A^B J(r_x,\varphi_x) e^{-\int_x^{B} K(r_{x'},\varphi_{x'})\,dx'}\,dx \,\, ,
\end{equation}
where $I_{\rm BC}$ is the boundary condition value. 
If point $A$ is on the inner boundary, then $I_{\rm BC}$ is the inner boundary condition, which is equation~(\ref{eq:opticlower}) as in the main text. 
If point $A$ is on the outer boundary, then $I_{\rm BC}$ is the outer boundary condition, which is zero.
The formal solution is calculated by numerical integration.

Note for a given $r$, the $\varphi$ parameter space is divided into two regions: one where point $A$ is on the inner boundary and one where it is on the outer boundary.
The dividing line $\varphi_{\rm div}$ is given by
\begin{equation}
    \varphi_{\rm div}=
    \arcsin\left(\frac{r_{\rm min}} r\right) 
\end{equation}
which always satisfies $0\leq\varphi_{\rm div}\leq\pi/2$. 
The specific intensities of the two regions are not continuous across the dividing line since the boundary condition $I_{\rm BC}$ changes across the dividing line. 
Therefore, we use two neural networks, one for each region.

\begin{figure}
    \centering
    \plotone{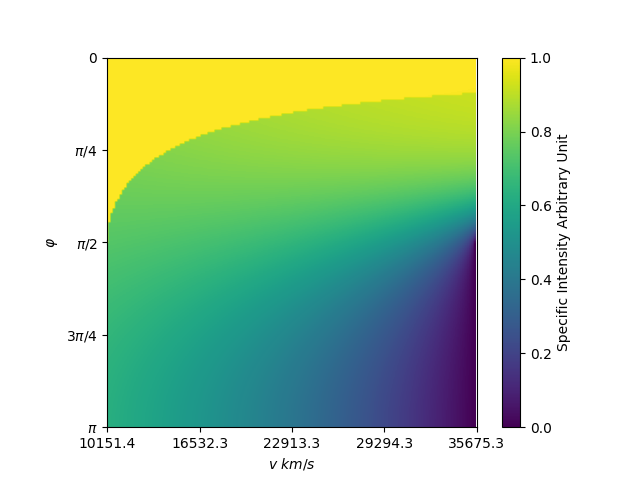}
    \caption{The specific intensity formal solution for our SN~Ia atmosphere with specific intensity as a function of velocity coordinate and viewing angle $\varphi$ given in color format for a single representative frequency with  $k_{\rm abs}$ and $j_{\rm em}$ set to constant values. 
    The specific intensity increases as color varies from purple to yellow. }
    \label{fig:illuTwoField}
\end{figure}

Figure \ref{fig:illuTwoField} shows an example of the formal solution. 
The discontinuity at the dividing line can clearly be seen as the curve separating the yellow color (which characterizes beams starting on the inner boundary) and the green and bluer colors (which characterize beams starting on the outer boundary). 

\end{document}